\RequirePackage{fix-cm}
\documentclass[twocolumn]{svjour3}          
\smartqed  

\newcommand{\pa}{\partial}

\newcommand{\be}{\begin{equation}}
\newcommand{\ee}{\end{equation}}

\usepackage{amsmath, amsfonts, amssymb, newlfont, graphicx}

\begin{document}

\title{Reduction of dimension for nonlinear dynamical systems}


\titlerunning{Reduction of dimension for nonlinear dynamical systems}        

\author{Heather A. Harrington and Robert A. Van Gorder}


\institute{H.A. Harrington \at
Mathematical Institute, University of Oxford, Andrew Wiles Building, Radcliffe Observatory Quarter, Woodstock Road, Oxford OX2 6GG United Kingdom \\
              \email{harrington@maths.ox.ac.uk} \\
R.A. Van Gorder \at
Mathematical Institute, University of Oxford, Andrew Wiles Building, Radcliffe Observatory Quarter, Woodstock Road, Oxford OX2 6GG United Kingdom \\
              \email{Robert.VanGorder@maths.ox.ac.uk}           
}

\date{Received: date / Accepted: date}

\maketitle

\begin{abstract}
We consider reduction of dimension for nonlinear dynamical systems. We demonstrate that in some cases, one can reduce a nonlinear system of equations into a single equation for one of the state variables, and this can be useful for computing the solution when using a variety of analytical approaches. In the case where this reduction is possible, we employ differential elimination to obtain the reduced system. While analytical, the approach is algorithmic, and is implemented in symbolic software such as {\sc MAPLE} or {\sc SageMath}. In other cases, the reduction cannot be performed strictly in terms of differential operators, and one obtains integro-differential operators, which may still be useful. In either case, one can use the reduced equation to both approximate solutions for the state variables and perform chaos diagnostics more efficiently than could be done for the original higher-dimensional system, as well as to construct Lyapunov functions which help in the large-time study of the state variables. A number of chaotic and hyperchaotic dynamical systems are used as examples in order to motivate the approach.
\keywords{reduction of dimension \and differential elimination \and nonlinear dynamics \and chaotic attractors \and computation of chaos}
\end{abstract}

\section{Introduction}
Nonlinear dynamical systems are ubiquitous in mathematics, engineering, and the sciences, with many real-world phenomenon governed by such nonlinear processes. In particular, non-equilibrium and chaotic dynamics are a continuing area of active research for applied mathematicians, as approximating such dynamics accurately and efficiently can be quite challenging. In the present paper, we shall consider reduction of dimension for nonlinear dynamical systems. This approach has previously been employed in the literature in order to enable the construction of Lyapunov functions \cite{MM1} and equilibrium dynamics \cite{MM2}, as well as to allow one to more easily approximate chaotic attractors analytically \cite{uc1,uc2,uc3,uc4}. One method for reduction of dimension is differential elimination, in which one algorithmically reduces the nonlinear dynamical system into a single ordinary differential equation (ODE) for one of the state variables. However, this is possible only when the system reduces to an ODE; if the reduction is instead to an integro-differential equation, the process is not algorithmic and specific cases must be handled with more individual care. Our focus shall be on dynamical systems giving chaotic dynamics, but the approach can certainly be applied for non-chaotic ODE systems. We give an overview of reduction of dimension, after which we demonstrate in several ways why one might wish to apply this technique.

With the wide range of numerical methods available for solving nonlinear first-order ODE systems of even high order, one may wonder why it might be advantageous to convert such systems into a single higher-order ODE. We shall mention several situations in which the differential elimination, and more generally reduction of dimension, may prove useful. We then outline the paper.

Often times, if one is trying to approximate the solution to a nonlinear system through some sort of analytical approximation, via series, perturbation, or more complicated approaches, one quickly finds that the coupled equations require balancing many terms coming from the expansion for each of the state variables. In the case of a single state variable governed by a higher-order ODE, one need only track terms in a single asymptotic expansion. This approach has been applied when using Taylor series, approximate Fourier series, and asymptotic expansions in other types of basis functions to the solution of a system of nonlinear ODE. In hybrid analytic-numeric methods, such as the homotopy analysis method \cite{ham1,ham2,ham3}, such reductions of a system to a single equation also simplifies the optimization problem which is solved to obtain the error-minimizing solution (see, for instance, \cite{MM2}, where the present approach is used in such a capacity). Therefore, the reduction of order can greatly reduce the complexity of analytical calculations under several frameworks. 

Contraction maps or Lyapunov functions are useful tools for discussing the convergence of solutions to nonlinear dynamical systems to large-time steady or quasi-steady dynamics. In situations where contraction maps or Lyapunov functions are known for a given dynamical system, the state variable governed by a single higher-order ODE necessarily results in a contraction map in the single state variable. However, as is well known to those studying stability of nonlinear systems, it is not often easy to obtain contraction maps for complicated systems. As we shall show here, it is possible to use the reduction of a system to a single higher-order ODE in order to construct a contraction map for the state variable governed by the aforementioned higher-order ODE. The existence of such a map can then be used to deduce the large-time dynamics of the state variable, as well as for the other state variables in the original system. One example of this is given in \cite{MM1}, and other examples are provided in Section 5.

Related to both the topic of analytical approximations and Lyapunov functions would be long time dynamics and equilibrium behavior of nonlinear dynamical systems. Indeed, in order to study the equilibrium structure of a high-order system of ODEs, one must solve a coupled system of nonlinear algebraic equations in order to recover the fixed points for the state variables. First reducing the system to a single ODE allows one to obtain a single nonlinear algebraic equation for the fixed point of a single state variable, which can then be used to recover the fixed points of the other state variables. Therefore, when such a reduction to a single ODE is possible, the need to solve a nonlinear algebraic system for all of the fixed points simultaneously is eliminated, resulting in what is often a far less computationally demanding problem.

Another topic is great recent interest in nonlinear science has been both the synchronization of chaos \cite{sync1,sync2,sync3} and the control of chaos. In situations where one is interested in mitigating the possibility of emergent chaos, one can couple a chaotic system to various control terms, or indeed to additional dynamical systems, which may lend a degree to stability. Under such approaches, one often increases the complexity or even the dimension of the dynamical system being solved. As such, methods to reduce the dimension of such systems could improve compatibility. Furthermore, since the control of chaos is often linked to a control term which itself is determined by a Lyapunov function, the construction of contraction maps through the reduction approach outlined here could be of great use.

As stated before \cite{CM1}, the competitive modes analysis gives an interesting link between the geometry of phase space possibly yielding chaotic trajectories (recall that the competitive modes requirements appear to be a necessary, albeit not sufficient, condition for chaos \cite{CM2,CM3,CM4,CM5,CM6,CM7}). Conversely, the differential elimination may cast light into the geometry of solutions in the space of derivations. Since this result of the differential elimination is a single higher-order ODE, and since any chaos emergent from the nonlinear system should be encoded in the single higher-order ODE, the differential algebraic structure of such equations may cast light into practical geometric tools by which one may study systems in which chaos is observed. In particular, through this reduction approach, the calculation of mode frequencies in the standard competitive modes analysis becomes much simpler. 

This paper is outlined as follows. In Section 2, we provide an algorithmic approach, to differential elimination for nonlinear dynamical systems based upon differential algebra. First laying out the general theory, we then give specific {\sc MAPLE} code for performing the differential elimination in a systematic manner. The algorithmic approach is useful in the case where the dynamical system can be reduced to a single ODE in terms of only one of the state variables. In Section 3 we implement the approach in order to reduce a variety of chaotic and hyperchaotic systems, finding that the form of the nonlinearity in the dynamical system will strongly influence the reducibility properties. However, in cases where the dynamical system is not reducible using differential elimination, one may still obtain more complicated reductions, for instance in terms of integrals, resulting in more complicated integro-differential equations for the reduced state variable. The possible results are illustrated through concrete examples for the R\"ossler system (which is completely reducible), the Lorenz system (which is partially reducible - that is, reducible in some but not all state variables), and the Qi-Chen-Du-Chen-Yuan (which is irreducible under differential elimination, but which can be reduced to an integro-differential equation). We give summarizing observations regarding the reducibility of dynamical systems in Section 4. 

The remainder of the paper is devoted to applications of reduction of dimension for dynamical systems. In Section 5, we demonstrate that reduction of dimension can be useful for obtaining contraction maps and Lyapunov functions, which in turn may be used to determine asymptotic stability of dynamical systems and also to control chaos in such systems. In Section 6 we demonstrate that reduction of dimension can be used to simplify calculations involved in certain techniques for studying the solutions of nonlinear dynamical systems. Indeed, when applicable, we find that the approach greatly reduces the number of nonlinear algebraic equations required to be solved when constructing trajectories in state space via undetermined coefficient methods by a factor of $1/n$, where $n$ is the dimension of the dynamical system, meaning that the number of equations needing to be solved will not increase with the size of the system. Furthermore, when applying the competitive modes analysis (which is a type of diagnostic criteria for finding chaotic trajectories in nonlinear dynamical systems), we find that only one binary comparison is needed of one first reduces the dimension of the dynamical system so that there is a single equation for one state variable. In contrast, there are normally of order $2^{n-1}$ comparisons needed for an $n$-dimensional dynamical system. In Section 7 we provide a discussion and possible avenues for future work.

\section{Algebraic approach to differential elimination}
Systems of differential equations are ubiquitous and widely studied. Ritt \cite{Ritt50} 
and Kolchin \cite{Kolch73} pioneered the field of differential algebra, an algebraic theory for studying solutions of ordinary and partial differential equations. We are particularly interested in differential elimination, an algorithmic subtheory that can simplify systems of parameterized algebraic differential equations. This permits one to reduce the dimension of a dynamical system so that one is left with a single ODE in the state variable. 

\subsection{Algebra preliminaries}
Here we briefly review concepts from algebra and differential algebra. For reference books in differential algebra, see \cite{Ritt50,Kolch73}. 
If $I$ is a subset of a ring $R$ then $(I)$ is the (algebraic) {\em ideal} generated by $I$. 
Let $I$ be an ideal of $R$. Then $\sqrt I$ denotes the {\em radical} of $I$.
A {\em derivation} over a ring $R$ is a map $ R \mapsto R$ which satisfies (we write $\dot{a}$ is the derivative of $a$), for every $a,b \in R$, $\dot{(a+b)}=\dot{a} + \dot{b}$ and $\dot{(a~b)} = (\dot{a})b+a(\dot{b}).$ 
The field of differential algebra is based on the concept of a {\em differential ring} (resp. field), which is a ring (resp. field) $R$ endowed with a set of derivations that commutes pairwise.  A {\em differential ideal} $[I]$ of a differential ring $R$ is an ideal of $R$ stable under the action of derivation.

Differential algebra is more similar to commutative algebra than analysis. In commutative algebra, Buchberger solved the membership problem (tests whether a given polynomial is contained in a given ideal) through the theory of Gr\"obner bases \cite{Buch65}. From algebraic geometry, we know the set of polynomials which vanish over the solutions of a given polynomial system form an ideal and even a radical ideal \cite{ZS58}. In the case of differential equations, the set of differential polynomials which vanish over the analytic solutions of a given system of differential polynomial equations form a differential ideal and even a radical differential ideal \cite{Ritt50}.
Ritt solved the theoretical problem (of membership for radical differential ideals) and developed algorithmic tools to solve systems of polynomial ODE and PDEs; however, Ritt's algorithm requires factorization. 

Due to the complexity of factorization, Boulier and co-authors avoided it by developing the Rosenfeld-Gr\"obner algorithm, based on the work of Seidenberg and Rosenfeld, and incorporating Gr\"obner bases \cite{Seid56,Ros59,BLOP95}. Since then, the algorithm has been improved both theoretically and practically \cite{BL00,BLOP09,HT14} and it no longer requires Gr\"obner bases. It is available in the {\tt DifferentialAlgebra} package in {\sc MAPLE} \cite{BLOP09} and {\tt SageMath} as an interface for the BLAD and BMI libraries \cite{blad,sage}.

Algorithmically, differential elimination involves manipulation of finite subsets of a differential polynomial ring $R = K\{U\}$ where $K$ is the differential field of coefficients (i.e. $K = \mathbb{Q}$), and $U$ is a finite set of dependent variables. The elements of $R$ are differential polynomials, which are polynomials built over the infinite set of all derivatives $\Theta U$, of the dependent variables. Consider a system $\Sigma$ of polynomial differential equations, here, we consider the Lorenz system of three ordinary differential equations:
\be\begin{aligned}
\dot{x_1} & = a(x_2-x_1)\,,\\
\dot{x_2} & = x_1(b-x_3)-x_2\,,\\
\dot{x_3} & = x_1x_2 - cx_3\,.
\end{aligned}\ee
The Lorenz system can be re-written as:
 \be \Sigma = \begin{cases}
-\dot{x_1} + a(x_2-x_3)=0\,,\\
-\dot{x_2} + x_1(b-x_3)-x_2=0\,,\\
-\dot{x_3} + x_1x_2 - cx_3=0\,.
\end{cases}\ee
The Rosenfeld-Gr\"obner algorithm takes as an input a differential system $\Sigma$ and a {\em ranking}. 
A ranking $>$ is any total ordering over the set $\Theta U$ of all derivatives of the elements of $U$ which satisfies the following axioms: $a<\dot{a}$ and $a<b \Rightarrow \dot{a}<\dot{b} $ for all $a,b \in \Theta U$. The Rosenfeld-Gr\"obner algorithm transforms $\Sigma$ into finitely many systems called regular differential systems, which reduces differential problems to purely algebraic ones that are triangular. The next step is purely algebraic and transforms the regular differential system into finitely many characteristic presentations, $C_1, \ldots C_r$. Rosenfeld-Gr\"obner outputs this finite family $C_1,\ldots C_r$ of finite subsets of $K\{U\} \setminus K$, where each $C_i$ defines a differential ideal $[C_i]$. The radical $\sqrt{ [\Sigma]}$ of the differential ideal generated by $\Sigma$ is the intersection presented by characteristic sets: $$\sqrt{ [\Sigma]} = [C_i]\cap \cdots \cap [C_r].$$  Note differential ideals $[C_i]$ do not need to be prime, however by Lazard's lemma, they are necessarily radical. Differential algebra elimination has proven useful for parameter estimation, identifiability, and model reduction of biological and chemical systems \cite{Mesh12,Boul07}.


\subsection{Computational Method}
We demonstrate reduction in dimension via differential elimination algorithm {\tt RosenfeldGroebner} in the {\tt DifferentialAlgebra} package implemented in {\sc MAPLE}. For sake of using a concrete example, we choose the Lorenz system.
First, we call the package:\\
{\tt with(DifferentialAlgebra):}\\
Next we input the Lorenz system:\\
{\tt sys := [-(diff(x1(t),t))+a*(x2(t)-x1(t)), \\-(diff(x2(t),t))+x1(t)*(b-x3(t))-x2(t), \\-(diff(x3(t),t))+x1(t)*x2(t)-c*x3(t)]}\\
Next we form our differential ring, embedding the rank of dependent variables in {\tt blocks} and independent variables in {\tt derivations}. Since we are considering {\em ordinary} differential equations, derivations is set to one ordering, time $t$. We remark that the {\tt DifferentialAlgebra} package enables differential elimination of PDEs by including additional inputs for the derivations (e.g.,\\
{\tt derivations=[u,x,t]}. Note, {\tt sys} is assumed to have coefficients in the field $\mathbb{Q}[x_1,x_2,x_3]$ obtained by adjoining the independent variables to the field of rationals and symbolic parameters $a,b,c$ are considered arbitrary in the coefficient field. To form the differential ring, we input:\\
{\tt R := DifferentialRing(blocks = [x3, x2, x1], \\
derivations = [t])}\\
Note that $x_1$ stands to the rightmost place on the list which identifies that we are attempting to reduce the differential equation to only one variable, i.e. $x_1(t)$. This ranking eliminates $x_3$ with respect to $x_2$, and then eliminates $x_2$ with respect to $x_1$. We now call the Rosenfeld Gr\"obner algorithm for our system and differential polynomial ring:\\
{\tt G := RosenfeldGroebner(sys, R)}\\
{\tt simplify(Equations(G[1], solved))}

This will return the characteristic presentation (which should be understood as an intersection), with the equations given by the ranking, with the final equation a single ODE for {\tt x1(t)}, provided that it exists and can be computed by the algorithm. In some cases the algorithm will keep running and therefore should eventually be terminated by the user. For such cases, it is unlikely that a reduction of the specified form exists. However, as we shall consider in the next section, when the reduction is to an integro-differential equation, rather than an ODE, the approach will not identify the reduced equation. 

\section{Reduction of Dimension: Applications}
Here we apply the method of differential elimination to several nonlinear dynamical systems known to give chaos, in order to see if these equations can be reduced. We first apply the algorithmic approach outlined in Section 2, finding that the approach gives a complete reduction (all state variables can be isolated and expressed as the solution to single uncoupled ODEs), a partial reduction (one or more, but not all, state variables can be isolated and expressed as the solution to single uncoupled ODEs), or returns no reduction (the algorithm does not complete in a fixed amount of time), in which case none of the state variables can be expressed as a solution to a single ODE reducible from the original system. For simplicity, we shall only consider autonomous systems. 

We consider a number of examples of chaotic systems in Table 1, with the results of the differential elimination algorithm given. We also give a summary of the dynamics of the example equations selected. Since the form of these equations may vary through the literature, we give a list of the specific form of the equations considered, in Appendix A. Note that we have considered the differential elimination algorithm for the arbitrary parameter values listed in Appendix A. Table 1 demonstrates that the structure of the dynamical system tends to play a strong role in whether the system can be reduced. Indeed, equations with a single nonlinearity tend to be completely or partially reducible, hence at least one state variable can be solved for via a single nonlinear ODE. On the other hand, the equations with many nonlinear terms or higher-order degree of nonlinearity (we consider only equations with polynomial nonlinearities) tend more often to be irreducible using the approach. Of the listed equations, note that the R\"ossler system is one of the few completely reducible systems, lending validity to the belief that it is indeed one of the simplest possible continuous- time dynamical systems giving chaos. Meanwhile, the commonly studied Lorenz system is only partially reducible under the approach. More complicated systems tend to be irreducible under the algorithm, and many of these give more complicated dynamics such as multiple scroll attractors. 

Note that the algorithm returns a `No' if a reduction is not obtained within a given time interval. For cases where the algorithm found a reduction, the computation time was fairly quick. We are therefore comfortable in assuming that a reduction to an ODE does not exist in cases where the the algorithm times out. For such cases, the system may still admit a reduction, but not strictly in derivatives of one of the state variables. One such example would be a system which is reducible to an integro-differential equation in one of the state variables, but never to simply an ODE.

\begin{table}\begin{center}
    \begin{tabular}{ | l | l | l |}
    \hline
    \textbf{System}  & \textbf{Dynamics} & \textbf{Reducible?}  \\ \hline
    Lorenz\cite{lor1,lor2} & 3D:1-2-2 & Partial\\ \hline
    Modified Chua's circuit\cite{modch1,modch2,modch3} & 3D:3-1-1 & Complete\\ \hline
	Chen-Lee\cite{cl1,cl2} & 3D:2-2-2 & Partial\\ \hline
    Rabinovich-Fabrikant\cite{rf1,rf2,rf3} & 3D:3-3-3 & No \\ \hline
    R\"ossler\cite{ros1,ros2} & 3D:1-1-2 & Complete \\ \hline
    Chen\cite{chen1,chen2} & 3D:1-2-2 & Partial \\ \hline
    L\"u\cite{lu1,lu2} & 3D:1-2-2 & Partial \\ \hline
    T-system\cite{t1,t2,t3} & 3D:1-2-2 & Partial \\ \hline
    Qi-Du-Chen-Chen-Yuan\cite{qii} & 4D:3-3-3-3 & No \\ \hline
    Qi-Chen-Du-Chen-Yuan\cite{qi} & 3D:2-2-2 & No \\ \hline
    Generalized Lorenz\cite{genlor1,genlor2,genlor3}& 3D:2-2-2 & No \\ \hline
    Blue-sky catastrophe\cite{bsc1,bsc2,bsc3} & 3D:3-3-3 & No \\ \hline
    Lorenz-Stenflo\cite{ls1,ls2,ls3} & 4D:1-2-2-1 & Partial \\ \hline
    Genesio-Tesi\cite{gt1,gt2} & 3D:1-1-2 & Partial \\ \hline
    Arneodo-Coullet-Tresser\cite{act} & 3D:1-1-3 & Partial \\ \hline
    \end{tabular}\end{center}
      \caption{List of chaotic systems and their reduction properties. The numbers in the dynamics column indicate the dimension of the system and then degree of each polynomial in the respective reaction functions. For instance, if $\dot{x}=A(x,y,z)$, $\dot{y}=B(x,y,z)$, $\dot{z}=C(x,y,z)$, then 3D:$\text{deg}(A)-\text{deg}(B)-\text{deg}(C)$ is reported, where $\text{deg}(A)$ denotes the degree of $A$, and so on. When the system has the property that it may be reduced to a single ODE in any state variable, we say that it is completely reducible, and record a `Complete'. If the system may be reduced to an ODE in one or more, but not all, state variables, we say the systems is partially reducible, and record a `Partial'. Finally, when a system is not reducible to a single ODE in any state variable, we record a `No'. We note that specific forms of some equations can change from paper to paper, so we record the specific equations used in Appendix A.}
\end{table}

We next consider hyperchaotic systems (chaotic systems giving two or more positive Lyapunov exponents) in Table 2. Again, we find that the more complicated the functional form of the nonlinarities, the less likely a system seems to be reducible. Furthermore, hyperchaotic generalizations of known chaotic systems appear to maintain their reducibility properties, since often a simple additional equation is added to make a chaotic system hyperchaotic. The hyperchaotic R\"ossler system is completely reducible, as was the related chaotic R\"ossler system, again suggesting that the chaotic and hyperchaotic R\"ossler systems are some of the simplest systems which still exhibit chaos and hyperchaos, respectively. 

\begin{table}\begin{center}
    \begin{tabular}{ | l | l | l |}
    \hline
    \textbf{System}  & \textbf{Dynamics} & \textbf{Reducible?}  \\ \hline
    R\"ossler \cite{hros1} & 4D:1-1-2-1 & Complete \\ \hline
    Chen \cite{hchen1,hchen2} & 4D:1-2-2-1 & Partial \\ \hline
    L\"u \cite{hlu} & 4D:1-2-2-2 & No \\ \hline
    Modified L\"u \cite{hmodlu} & 4D:2-2-2-1 & No \\ \hline
    Wang-Liu \cite{wl} & 4D:1-2-2-1 & Partial \\ \hline
    Jia \cite{jia1,jia2} & 4D:1-2-2-2 & Partial \\ \hline
    QWWC system \cite{qvvc1,qvvc2} & 4D:2-2-2-2 & No \\ \hline
    \end{tabular}\end{center}
      \caption{List of hyperchaotic systems and their reduction properties. The labeling is the same as was given in Table 1. We note that specific forms of some equations can change from paper to paper, so we record the specific equations used in Appendix B.}
\end{table}

The results indicate that completely reducible systems are perhaps the simplest systems giving chaos or hyperchaos. Again, this would support the qualitative and topological claims that the R\"ossler systems are some of the simplest possible equations permitting chaos \cite{rossTop}, as they each involve only a single quadratic nonlinearity. On the other hand, systems with stronger polynomial nonlinearities, or systems with many nonlinear terms, appear to often be irreducible under differential elimination. Note that for cases where the reduction might involve integrals, resulting in a type of integro-differential equation, the differential elimination algorithm would miss such a reduction, even though it exists. This is due to the fact that the differential elimination algorithm is working over the ring of derivations, which does not include integrals. Indeed, since integral operators are fairly cumbersome to introduce compared to their differential operator counterparts (we discuss this later in Section 7), obtaining an algorithmic approach including integrals would be challenging. Therefore, the differential elimination algorithm outlined in Section 2 appears to be a very useful tool for reducing the dimension of dynamical systems, provided that a reduction to a single ODE exists. For the more complicated models, we find the need to proceed on a case-by-case basis with manual manipulations due to any integration needed. 

We demonstrate reduction of dimension for chaotic systems into single higher-order ODEs in the next three subsections. We pick a case where all state variables can be isolated (the R\"ossler system), a case where one of the state variables can be isolated in terms of a differential equation (the Lorenz system), and finally a case where none of the state variables can be isolated in terms of a differential equation (the Qi-Chen-Du-Chen-Yuan system) so that any reduction would necessarily involve integrals. For all cases considered, we let $x,y,z\in C^n(\mathbb{R})$ where $n$ is the dimension of the relevant dynamical system, and we take $a,b,c \in \mathbb{R}$ to be parameters.

\subsection{R\"ossler System}
The R\"ossler equation \cite{ros1,ros2} reads 
\be\begin{aligned} \label{Rossler}
\dot{x} & = -y-z\,,\\
\dot{y} & = x + ay\,,\\
\dot{z} & = b + x(x-c)\,.
\end{aligned}\ee

We first obtain the ODE for $y(t)$. Note from the second equation that $x=\dot{y}-ay$, so that $\dot{y}=\ddot{y}-a\dot{y}$ and hence from the first equation we have $z=-\ddot{y}+a\dot{y}-y$. Placing these into the third equation, and performing algebraic manipulations, we obtain
\be \label{RossY}
\dddot{y} - a\ddot{y} + \dot{y} - \left(\ddot{y} - a\dot{y}+y\right)\left( \dot{y} -ay -c \right) +b =0\,.
\ee
Note that this equation is third order, and therefore the information of the three-dimensional system \eqref{Rossler} can be encoded in this single ODE. By similar manipulations, one may arrive at an equation for $x(t)$,
\be 
\left(a+c-x\right)^2\left\lbrace \left(\frac{d}{dt} -(x-c)\right)\frac{\ddot{x}-a\dot{x}+x+b}{a+c-x} -b \right\rbrace =0\,,
\ee
and an equation for $z(t)$, 
\be 
z^3 \left( \frac{d^2}{dt^2} - a\frac{d}{dt} +c \right)\frac{\dot{z}-b}{z} + z^3\dot{z} - az^4 + cz^3 =0\,.
\ee

\subsection{Lorenz System}
The Lorenz system \cite{lor1,lor2} is given by
\be\begin{aligned} 
\dot{x} & = a(y-x)\,,\\
\dot{y} & = x(b-z) - y\,,\\
\dot{z} & = xy - cz\,.
\end{aligned}\ee
Observing from the first two equations that
\be 
y = x + \frac{1}{a}\dot{x}
\ee
and
\be 
z = b - \frac{\ddot{x}+(1+a)\dot{x}+x}{ax}\,,
\ee
the third equation can be used to obtain a single ODE for the state variable $x(t)$, viz.,
\be 
x^2\left( \frac{d}{dt} +c\right) \frac{\ddot{x}+(1+a)\dot{x}+x}{x}  +ax^4 +x^3\dot{x} -abcx^2 =0\,.
\ee
This agrees with what one obtains from the differential elimination. On the other hand, we observe that the algorithmic approach to differential elimination is useful for situations in which there is no obvious route to reduce a system into a single equation (through eliminations and substitutions). A good example of this is found when trying to obtain a differential equation for the state variable $z(t)$ alone. Using the differential elimination, we arrive at a rather complicated equation of the form
\be 
(b-z)(\dddot{z})^2 + P_1(z,\dot{z},\ddot{z})\dddot{z} + P_2(z,\dot{z},\ddot{z})=0\,,
\ee
where $P_1$ and $P_2$ are complicated polynomials that we do not list for sake of brevity. Interestingly, this is a fully nonlinear equation, since the highest order derivative enters into the equation nonlinearly. In contrast, the equation obtained for the state variable $x(t)$ is quasi-linear, since it is linear in the highest derivative. One could differentiate the equation for $z(t)$ in order to isolate the highest derivative, but by doing so one would increase the differential order of the system, thereby decreasing the regularity of the system. This is particularly important in cases where the solution $z(t)$ may only be $C^3(\mathbb{R})$.

When a system is nonlinear, there may of course be forms of the nonlinearity which do not permit one to obtain an equation for a single state variable in terms of that state variable and its derivatives. A good example of this is the state variable $y(t)$ in the Lorenz system. The algorithmic differential elimination finds no closed differential equation for $y(t)$. As it turns out, the reason for this is that any equation governing $y(t)$ alone will necessarily involve integral terms which cannot be eliminated (due to the nonlinearity of the equation). To see this, note that if we consider the first equation in the Lorenz system, which may be written in the form $(e^{at}x)' = ae^{at}y$, we find
\be 
x(t) = x_0 + ae^{-at}\int_0^t e^{as}y(s)ds\,.
\ee
Here $x_0$ is the initial value of the state $x(t)$, that is $x(0)=x_0$. Yet, from the second equation in the Lorenz system, we have $z = b - (\dot{y}+y)/x$, which yields
\be 
z(t) = b - \frac{\dot{y}+y}{x_0 + ae^{-at}\int_0^t e^{as}y(s)ds}\,.
\ee
Placing the representations for $x(t)$ and $z(t)$ into the third equation of the Lorenz system, and performing algebraic manipulations to simplify the resulting expression, we obtain
\be \begin{aligned}\label{lor_y}
& (\ddot{y}+(1+c)\dot{y}+cy)\left( x_0 + ae^{-at}\int_0^t e^{as}y(s)ds \right)\\
& \quad + (\dot{y}+y)\left( ay - a^2 e^{-at}\int_0^t e^{as}y(s)ds \right)\\
& \quad + y\left( x_0 + ae^{-at}\int_0^t e^{as}y(s)ds \right)^3\\
& \quad -cb \left( x_0 + ae^{-at}\int_0^t e^{as}y(s)ds \right) =0\,.
\end{aligned}\ee 
Note that the equation both involves an integral and is non-autonomous.

\subsection{Qi-Chen-Du-Chen-Yuan System}
We now consider the Qi-Chen-Du-Chen-Yuan (QCDCY) system \cite{qi}, which is given by 
\be\begin{aligned} 
\dot{x} & = a(y-x) +yz\,,\\
\dot{y} & = bx - y - xz \,,\\
\dot{z} & = xy - cz \,.
\end{aligned}\ee
The differential elimination algorithm indicates there is no reduction to a single ODE in any of the three state variables. This system has a quadratic nonlinearity in each equation, and this added complication is behind the difficulties in obtaining such a reduction. However, we may still obtain an equation for a single state variable, if we are willing to include integral terms. Due to the complexity in obtaining such an equation, we shall restrict our attention to finding a single equation for the state variable $z(t)$, noting that similar approaches can be used to find a single equation for either of the other two state variables, $x(t)$ or $y(t)$. 

Let us begin by noting that the second equation in the QCDCY system implies $(e^t y)' = e^t (b-z)x$, while placing this into the third equation in the QCDCY system gives $(a+z)(\dot{z}+cz)=xe^{-at}(e^{at}x)' = x(\dot{x}+ax)$. This, in turn, implies that state variables $x(t)$ and $z(t)$ satisfy
\be 
e^{2at}(x(t))^2 = x_0^2 +2\int_0^t e^{2as}(a+z(s))(\dot{z}(s)+cz(s))ds\,.
\ee
where $x(0)=x_0$. The first equation in the QCDCY system has not been used, and we place this relation into that equation to obtain a single equation for the state variable $z(t)$. After several algebraic and differential manipulations, we arrive at the single equation
\be \begin{aligned}
& 2e^{2at}\left( 1-a-\frac{\dot{z}}{a+z} \right)(\dot{z}+cz)(a+z)\left(x_0^2 +J[z,\dot{z}]\right)\\
& \qquad +2\left(x_0^2 +J[z,\dot{z}]\right)\frac{d}{dt}\left(e^{2at}(\dot{z}+cz)(a+z)\right)\\
& \qquad -2e^{4at}(\dot{z}+cz)^2(a+z)^2 \\
& \qquad - 2(b-z)(a+z)\left(x_0^2 +J[z,\dot{z}]s\right)^2 = 0\,,
\end{aligned}\ee
where we have defined the integral operator
\be \label{J}
J[z,\dot{z}] = 2\int_0^t e^{2as}(a+z(s))(\dot{z}(s)+cz(s))ds\,.
\ee

Similar results can be obtained for the other state variables. The fact that the obtained equations involve an integral operator which cannot simply be differentiated away demonstrates why the differential elimination algorithm was not useful for this case. Still, performing the manipulations by hand, we have reduced the fairly complicated QCDCY system into a single integro-differential equation, thereby reducing the dimension of the original system.

\section{Reductions of $n$-dimensional dynamical systems}
We now give some summarizing remarks based on what we have seen in the previous sections. We shall assume that each system is coupled through at least one state variable (otherwise the state variables naturally separate into distinct lower-dimension equations, and the approach is not needed).

\subsection{Linear Systems}
For first order linear systems of dimension $n$, there is always a reduction into a single higher-order ODE. This follows from the process of Gaussian elimination. In the case that the matrix of coefficients for such a first-order system is full rank, the resulting higher-order ODE will be of order $n$. If the matrix of coefficients is singular, then the resulting higher-order ODE will be of order less than $n$. 

\subsection{Reducible Nonlinear Systems}
For first order nonlinear systems of dimension $n$, there are multiple possibilities, owing to the structure of the nonlinearity.

In cases where the system permits the complete differential elimination (an example being the Rossler equation), all state variables in a first order nonlinear system can be expressed in terms of a higher-order ODE. Note, however, that it is possible for the order of the single ODE to be different from the dimension $n$ of the first order system. As an example of this point, consider the system
\be\begin{aligned} 
\dot{x} & = x - y - z\,,\\
\dot{y} & = x^2 \,,\\
\dot{z} & = x - x^3 \,.
\end{aligned}\ee
Clearly, differentiation of the first equation gives $\ddot{x}= \dot{x} - \dot{y} -\dot{z} = \dot{x} - x^2 +x -x^3$. So, we obtain
\be 
\ddot{x} - \dot{x} - x + x^2 + x^3 =0\,,
\ee
which is a second order equation for the state variable $x(t)$, even though the original system was first order. A similar example can be found in \cite{MM1}, where a fourth order nonlinear dynamical system was reduced to a single second order nonlinear ODE.

It is possible for a system to be reduced to a single equation, which is not an ODE. This was evident even for the Lorenz equation, where an equation for one of the state variables involves an integral term in addition to derivative terms. Note that the equation was not closed under any number of differentiations, due to the form of the intergral terms. As such, the single reduced equation for the state variable could never be expressed strictly as an ODE of any finite order. Note that this can occur for one of the state variables, while a different state variable might satisfy a finite order ODE. For such cases, the nonlinearity in the system results in their being certain favored state variables with which to perform the reduction to a single ODE.

\subsection{Differentially Irreducible Nonlinear Systems}
We have observed that for more complicated nonlinear dynamical systems, there is no reduction to a single ODE in one state variable. While it may be the case that differential elimination does not pick up an ODE that does exist, it seems as though the failure of differential elimination is a sign that integrations will be needed in order to reduce the dimension of such systems. Indeed, when integrations of this kind are called for, the manipulations are no longer confined to the specified differential ring, and the differential elimination cannot be performed. While one can attempt these integrations manually, as opposed to algorithmically, obviously it would be desirable to have some kind of algorithmic approach. Perhaps one may adjoin integral operators to the differential ring, in order to perform reductions for more complicated nonlinear systems. This would likely work in cases like the Lorenz system, for which there is partial reducibility under differential elimination. For instance, if one were to define a new variable $Y(t) = \int_0^t e^{as}y(s)ds$, then one would obtain a non-autonomous ODE for $Y(t)$ from equation \eqref{lor_y}. Therefore, this fairly simple integral transformation, in addition to differential operators, can reduce the dimension of the Lorenz system with respect to the state variable $y(t)$. However, in cases like that of the QCDCY system, note that the form of the integral operator given in \eqref{J} is rather complicated, depending nonlinearly on both the state variable $z(t)$ and its derivative $\dot{z}(t)$. For such cases, there is no combination of elementary integral transforms that can be adjoined to the differential ring which would permit reduction of dimension to a single ODE. As such, it appears as though reduction of dimension for certain more complicated systems will result in reductions to integro-differential equations, rather than ODEs, for some fundamental reason related to how complicated the original dynamical system is. Therefore, the study of possible algorithmic methods for the reduction of dimension for dynamical systems into single scalar integro-differential equations would be an interesting and potentially very useful area of future work.

\section{Contraction Maps and Lyapunov Functions}
Turning out attention now to practical applications for reduction of dimension, recall that contraction maps and Lyapunov functions are useful tools for studying the asymptotic stability of nonlinear dynamical systems. In this section, we use the three examples worked explicitly in Section 3 in order to demonstrate the utility of reduction of dimension for finding Lyapunov functions. Using these results, we can recover stability results for these dynamical systems which were obtained through other approaches, and which agree with existing results in the literature. 

\subsection{R\"ossler System}
The R\"ossler system has two equilibrium values, $\pm y^*$, for $y(t)$, and the constant $y^*$ must satisfy the quadratic equation 
\be \label{algeb}
a (y^*)^2 + cy^* + b =0\,.
\ee

In order to discover a Lyapunov function for the R\"ossler system, it is tempting to assume a bowl-shaped map of the form $\alpha x^2 + \beta y^2 + \gamma z^2$, or minor variations on this theme involving higher power polynomials of even order, but the approach evidently proves fruitless. Therefore, we shall use one of the three equations obtained for the isolated state variables of the R\"ossler system.

Consider equation \eqref{RossY} for the R\"ossler system \eqref{Rossler}. Let us write $Y(t) = y(t)-y^*$ in the neighborhood of either equilibrium value $y^*$. This transformation will prove useful, as the Lyapunov function needs to vanish at the equilibrium value selected. (There is therefore the need to construct such a function in a neighborhood of each equilibrium point.) Under this transformation, \eqref{RossY} is put into the form
\be \label{Ross_why}
\dddot{Y} - a\ddot{Y} + \dot{Y} - \left(\ddot{Y} - a\dot{Y}+Y\right)\left( \dot{Y} -aY \right)  =0\,.
\ee 
Let us define a function $m=\ddot{Y} - a\dot{Y}+Y$ so that \eqref{Ross_why} is put into the form
\be \label{Ross_m}
\dot{m} - (\dot{Y}-aY)m =0\,.
\ee
Observe that \eqref{Ross_m} can be written as
\be 
\dot{m} - e^{at}\left(e^{-at}Y\right)' m =0\,.
\ee
From this, we recover
\be 
\ddot{Y} - a\dot{Y}+Y = m = m_0\exp\left( \int_0^t e^{a\zeta}(e^{-a\zeta}Y(\zeta))' d\zeta \right)\,,
\ee
where $m_0$ is a constant of integration. As we are interested in recovering information about the asymptotic stability of the R\"ossler system, let us pick the initial condition $Y(0)=\epsilon$. This corresponds to setting the initial condition such that it is contained within a neighborhood of the equilibrium value. Let us also restrict $|a|<2$ (this will simplify the mathematics, and is consistent with the physics of the R\"ossler system). Then, we obtain
\be \begin{aligned}
Y(t) & = \epsilon e^{at/2}\left\lbrace \cos\left(\frac{\sqrt{4-a^2}}{2}t\right) + C \sin\left(\frac{\sqrt{4-a^2}}{2}t\right)\right\rbrace\\
& \quad + m_0 \int_0^t K(t,s)\exp\left( \int_0^s e^{a\zeta}(e^{-a\zeta}Y(\zeta))' d\zeta \right) ds\,,
\end{aligned}\ee
where $C$ is a constant that will depend on the initial value of $\dot{Y}(0)$ (the value of which will not impact our analysis) and $K(t,s)$ is the kernel
\be 
K(t,s) = e^{\frac{a}{2}(t-s)}\sin\left(\frac{4-a^2}{2}(t-s)\right)\,.
\ee
Observe that for $-2 < a < 0$ the map is a contraction. Given arbitrarily small $\epsilon>0$, for large enough time $\tilde{t}(\epsilon)>0$, the solution $Y(t)$ will lie in a neighborhood $-\epsilon < Y(t) < \epsilon$ for all $t>\tilde{t}(\epsilon)$. Therefore, $Y\rightarrow 0$ as $t\rightarrow \infty$. Yet, by definition of $Y(t)$, this implies $y \rightarrow y^*$ as $t\rightarrow \infty$. Using this, one may shown that $x\rightarrow -ay^*$ and $z\rightarrow -y^*$ as $t\rightarrow \infty$. Hence, we have shown that $a<0$ gives a stable solution, which was already known from different work. The nice thing about this approach is that is allows us to bypass a linear stability analysis involving the calculation of eigenvalues at the algebraic solution to $y^*$ found from \eqref{algeb}. Indeed, we did not even need to calculate the equilibrium value $y^*$ for the present analysis, as the analysis holds for an arbitrary equilibrium value satisfying \eqref{algeb}.

\subsection{Lorenz System}
In order to find a Lyapunov function for the Lorenz system in a neighborhood of the zero equilibrium $(x,y,z)=(0,0,0)$, let us assume a bowl type function of the form
\be 
V(x,y,z) = \alpha x^2 + \beta y^2 + \gamma z^2\,,
\ee
where $\alpha>0$, $\beta >0$, and $\gamma >0$ are constant parameters to be selected. recall that physically interesting model parameters $a$, $b$, and $c$ are positive. Then, the time derivative of $V$ is given by
\be 
\frac{1}{2}\dot{V} = -\alpha a x^2 - \beta y^2 - \gamma c z^2 + (\alpha a + \beta b)xy + (\gamma -\beta b)xyz\,.
\ee
Clearly, we should take $\gamma = \beta b$. Note that
\be 
-(\sqrt{\alpha a}x - \sqrt{\beta}y)^2 = -\alpha a x^2 - \beta y^2 + 2\sqrt{\alpha\beta a}xy\,.
\ee
Then, 
\be 
\frac{1}{2}\dot{V} = -(\sqrt{\alpha a}x - \sqrt{\beta}y)^2 - \gamma c z^2 + (\alpha a + \beta b - 2\sqrt{\alpha\beta a})xy \,,
\ee
hence $\dot{V} \leq 0$ provided that $\beta b < 2\sqrt{\alpha\beta a} - \alpha a$ (since this would imply $-\alpha a x^2 - \beta y^2 + (\alpha a + \beta b)xy <0$). Let us pick $\beta = \alpha a$. Then, the condition reduces to $b<1$. As $\alpha >0$ was arbitrary, we set $\alpha = \frac{1}{2}$. This means that whenever $a>0$, $0<b<1$, and $c>0$, there exists a Lyapunov function 
\be 
V(x,y,z) = \frac{1}{2} x^2 + \frac{a}{2} y^2 + \frac{ab}{2} z^2\,,
\ee
since $V(0,0,0)=0$, $|V|\rightarrow \infty$ as $|(x,y,z)|\rightarrow \infty$ (radially unbounded), and $\dot{V} <0$ for $(x,y,z)\neq (0,0,0)$. Interestingly, the condition $<b<1$ is exactly the stability condition known in the literature \cite{lor2}. Therefore, parameters implying the existence of this contraction map correspond to known stable parameters.

Now, if we were to seek such a map for only one of the state variables, then using what we have obtained in Section 3, we find that there exists a contraction map 
\be \begin{aligned}
\hat{V}(x,\dot{x},\ddot{x}) & = \frac{1}{2} x^2 + \frac{a}{2} \left( x + \frac{\dot{x}}{a} \right)^2 \\
& \qquad + \frac{ab}{2} \left( b - \frac{\ddot{x}+(1+a)\dot{x} +ax}{ax} \right)^2
\end{aligned}\ee
for the state variable $x(t)$. Then, one may verify $\dot{\hat{V}}<0$ away from the equilibrium $x=0$. One can obtain similar contraction maps in either of the other two state variables.

\subsection{Qi-Chen-Du-Chen-Yuan System}
In order to find a contraction map for the Qi-Chen-Du-Chen-Yuan (QCDCY) system, we begin with the bowl shaped assumption for a Lyapunov function about the equilibrium $(x,y,z)=(0,0,0)$,
\be 
V(x,y,z) = \alpha x^2 + \beta y^2 + \gamma z^2\,,
\ee
where $\alpha>0$, $\beta >0$, and $\gamma >0$ are constant parameters to be selected. Differentiating with respect to $t$ and using the three constituent equations of the QCDCY system, we have
\be 
\dot{V} = (\alpha - \beta +\gamma)xyz + (\alpha a + \beta b)xy - \alpha a x^2 - \beta y^2 - \gamma c z^2 \,.
\ee
Since we need $\alpha>0$, $\beta >0$, and $\gamma >0$, we should consider model parameters satisfying $a>0$ and $c>0$. To remove the first term, which is hyperbolic in nature, we should choose $\beta = \alpha + \gamma$. Meanwhile, the remove the second term, which is also hyperbolic, we should set $\beta = -\frac{a}{b}\alpha$ for non-zero $b$. Since all other parameters are positive, we must require $b<0$. Then, $\beta = \frac{a}{|b|}\alpha$, and placing this into $\beta = \alpha + \gamma$ gives $\gamma = \frac{a-|b|}{|b|}$. As we need $\gamma >0$, this gives the added restriction $a > |b|$. The parameter $\alpha$ is arbitrary, so we take $\alpha = \frac{1}{2}$. We therefore obtain
\be 
V(x,y,z) = \frac{1}{2} x^2 + \frac{a}{2|b|} y^2 + \frac{a-|b|}{2|b|} z^2\,,
\ee
and this candidate function is indeed a contraction map satisfying $V(0,0,0)=0$, $\dot{V}<0$ for $(x,y,z)\neq (0,0,0)$, and $|V|\rightarrow \infty$ as $|(x,y,z)|\rightarrow\infty$, provided that the parameter restrictions $a>|b|$, $b<0$, and $c>0$ hold. Therefore, under these parameter restrictions, the zero equilibrium is asymptotically stable for the QCDCY system.

When we obtain a single equation for a state variable, even one containing integrals, we may similarly obtain a contraction map. Since we have obtained an equation for the state variable $z(t)$ in the QCDCY system in Section 3, we shall choose to construct a contraction map for that state variable here. Doing so, we find that
\be \begin{aligned}
&\hat{V}(z,\dot{z})  = \frac{a}{|b|}\frac{e^{2at} (\dot{z}+cz)^2}{x_0^2 +2\int_0^t e^{2as}(a+z(s))(\dot{z}(s)+cz(s))ds}\\
& \quad + e^{-2at}\left(x_0^2 +2\int_0^t e^{2as}(a+z(s))(\dot{z}(s)+cz(s))ds\right)\\
& \quad + \frac{a-|b|}{|b|}z^2
\end{aligned}\ee
satisfies $\dot{\hat{V}}<0$ for all $z\neq 0$, given that $a>|b|$, $b<0$, and $c>0$. Hence, $\hat{V}(z,\dot{z})$ is a contraction map for the state variable $z(t)$ when $a>|b|$, $b<0$, and $c>0$. With this, we have determined the stability of the zero equilibrium for the QCDCY system.

\section{Computational considerations for chaotic trajectories}
There are a variety of methods available for trying to find chaotic trajectories in nonlinear dynamical systems, and the approach highlighted in this paper does not add to collection of tools, explicitly. However, the reduction of dimension approach outlined in Section 2 can be used to make finding chaos in dynamical systems more efficient. To demonstrate this, we shall consider two rather distinct approaches, namely, the undetermined coefficients method for obtaining chaotic trajectories and the competitive modes analysis for identification of chaotic parameter regimes. For each of these approaches, we show that an application of reduction of dimension results in a simplification of each test for chaos. 

\subsection{Calculation of trajectories via undetermined coefficients}
When attempting to analytically calculate chaotic trajectories, even in an approximate sense, one often reduces the dimension of the governing equations. The reason for this lies in the fact that it is easier to consider an expansion for one state variable, rather than multiple state variables. To best illustrate this point, let us return to the R\"ossler equation \eqref{Rossler}. 

One popular method for approximating trajectories of chaotic systems analytically is the undetermined coefficient method \cite{uc1,uc2,uc3,uc4}. Since Taylor series expansions for nonlinear systems often have a finite region of convergence centered at the origin, yet the chaotic dynamics remain bounded in space, one often considers non-polynomial base functions. One popular choice would be a function of the form 
\be 
S(t;\left\lbrace A_j \right\rbrace_{j=-\infty}^{j = \infty},\alpha) = \begin{cases}
\sum_{j=0}^\infty A_{j} e^{-\alpha j t} & \text{for} ~ t\geq 0\,,\\
\sum_{j=0}^\infty A_{-j} e^{\alpha j t} & \text{for} ~ t< 0\,.
\end{cases}
\ee
In this expression, the $A_j \in \mathbb{R}$ and the parameter $\alpha >0$ are undetermined parameters which one typically will obtain in an iterative manner. Assuming such an expansion in time, it males sense to consider a solution the the R\"ossler system \eqref{Rossler} of the form $x(t) = S(t;\left\lbrace A_j \right\rbrace_{j=-\infty}^{j = \infty},\alpha)$, $y(t)=S(t;\left\lbrace B_j \right\rbrace_{j=-\infty}^{j = \infty},\alpha)$, and $z(t)=S(t;\left\lbrace C_j \right\rbrace_{j=-\infty}^{j = \infty},\alpha)$. Placing these equations into \eqref{Rossler}, one would obtain an infinite system of nonlinear algebraic equations for all of the coefficients and the temporal scaling $\alpha >0$. In practice, one would truncate these expansions, taking the sum over $-J \leq j \leq J$ for some $J>0$. As the solutions may converge slowly - if they converge at all (owing to the nonlinearity), one would need to solve $6J+1$ nonlinear algebraic equations. 

Assume, instead, that we wish to solve \eqref{RossY} by the approach described above. We would then insert the expansion for $y(t)$ into \eqref{RossY}. Assuming that we can solve the resulting nonlinear algebraic equations for the constants $\left\lbrace B_j \right\rbrace_{j=-\infty}^{j = \infty}$ and $\alpha$, we can then recover $x(t)$ and $z(t)$ by recalling $x = \dot{y}-ay$ and $z = -\ddot{y}+a\dot{y}-y$. From these expressions, it is simple to show $A_n = -(\alpha |n|+a)B_n$ and $C_n = -(\alpha^2 n^2 +a \alpha |n| +1)B_n$ for all $n\in \mathbb{Z}$. If we were to truncate the expansion for $y(t)$, in the manner described above, we would need to solve $2J+1$ nonlinear algebraic equations for $\left\lbrace B_j \right\rbrace_{j=-J}^{j = J}$ and $\alpha$, while the coefficients for $x(t)$ and $z(t)$ are immediately found once we know these parameters. This means that by first reducing the dimension of the ODE system, we would be able to reduce the computational complexity of the problem by a factor of three. For higher-dimensional system, the reduction in computational complexity will scale as the dimension of the system itself. In other words, a solution in term of the undetermined coefficient method will not depend on the size of the dynamical system provided that the dynamical system can be reduced in dimension to a single equation governing one state variable.

\subsection{Competitive modes analysis: A check for chaos}
The method of competitive modes involves recasting a dynamical system as a coupled system of oscillators \cite{CM1,CM2,CM3,CM4,CM5,CM6}. Consider the general nonlinear autonomous system of dimension $n$ given by
\begin{equation}
\dot{x_{i}}=f_i(x_1,x_2,...,x_n)\,. \label{autonomoussystem}
\end{equation} 
Differentiation of \eqref{autonomoussystem} once gives a coupled system of second order equations,
\begin{equation}\begin{aligned}
\ddot{x_i} =  \sum\limits_{j=1}^{n}f_j\frac{\partial f_i}{\partial x_j} = & -x_ig_i(x_1,x_2,...,x_i,...,x_n)\\
&  +h_i(x_1,x_2,...,x_{i-1},x_{i+1},...,x_n)\label{CM}\,.
\end{aligned}\end{equation}
When a $g_i$ is positive, its respective $i$th equation behaves like an oscillator. The following conjecture is posed in \cite{CM2}:\\
\\
\noindent {\bf Competitive Modes Requirements}: The conditions for dynamical systems to be chaotic are given by:\\
(A) there exist at least two modes, labeled $g_i$ in the system;\\
(B) at least two $g$'s are competitive or nearly competitive, that is, for some $i$ and $j$, $g_i \approx g_j >0$ at some $t$;\\
(C) at least one of the $g$'s is a function of evolution variables such as $t$; and\\
(D) at least one of the $h$'s is a function of system variables.\\

The requirements (A)-(D) essentially tell us that a condition for chaos is that two or more equations in \eqref{CM} behave as oscillators ($g_i>0$), and that two of these oscillators lock frequencies at one or more times. In practice, we find that the frequencies agree at a countably infinite collection of time values \cite{CM1,CM6}. The frequencies should be functions of time (i.e., we have nonlinear frequencies), and there should be at least one forcing function which depends on a state variable.

In order to consider all possible chaotic dynamics, one would have to compare each pair $g_i = g_j$, $i\neq j$, $i,j = 1,2,3,\dots ,n$. Accounting for symmetry, this gives $2^{n-1} -1$ matchings to consider. For high-dimensional dynamical systems, this number becomes rather large. As an example, in the case of a ten-dimensional system, there will be 511 possible matchings to be checked in order to ensure one has determined the possible chaotic regimes. For such situations, the approach is not particularly efficient, and a competitive modes analysis is often considered for systems of dimension three or four in the literature. 

Let us consider dynamical systems \eqref{autonomoussystem} which can be put into the form of a single equation for one state variable. As such an equation will encode the dynamics of the complete system, it is sufficient to consider a competitive modes analysis for the resulting equation. Suppose that the resulting equation has maximal derivative of order $p>0$ (where $p$ need not be equal to $n$, as we have seen in earlier sections). Then, associating $y(t)$ to this single state variable, we have
\be \label{y}
\frac{d^p y}{dt^p} = F\left( y, \frac{dy}{dt}, \dots , \frac{d^{p-1}y}{dt^{p-1}} \right)\,.
\ee
Since the competitive modes analysis relies on us obtaining a system of oscillator equations, let take $y=y_1$ and write the equation \eqref{y} as the system
\be \label{yy}
\dot{y}_1 = y_2\,\dots, \dot{y}_{p-1}=y_p\,, \dot{y}_p = F(y_1,\dots, y_p)\,.
\ee
Differentiation of \eqref{yy} results in the system of second order equations given by
\be \begin{aligned}
\ddot{y}_1 & = y_3\,,\\
& \vdots \\
\ddot{y}_{p-2} & = y_p\,,\\
\ddot{y}_{p-1} & = \dot{y}_p = F(y_1,\dots, y_p)\,,\\
\ddot{y}_p & = \sum_{i=1}^p \frac{\pa F}{\pa y_{i}}\dot{y}_i = \sum_{i=1}^{p-1} \frac{\pa F}{\pa y_{i}}y_{i+1} + \frac{\pa F}{\pa y_{p}} F(y_1,\dots, y_p)\,.
\end{aligned}\ee
The right hand side of the first $p-2$ second order equations do not depend on the state variable for each respective equation, so $g_1=\cdots = g_{p-2} =0$. Hence, these equations are never oscillators. Meanwhile, we can decompose the right hand sides of the latter two equations, so that
\be 
F(y_1,\dots, y_p) = - y_{p-1}g_{p-1} + h_{p-1}
\ee
and 
\be 
\sum_{i=1}^{p-1} \frac{\pa F}{\pa y_{i}}y_{i+1} + \frac{\pa F}{\pa y_{p}} F(y_1,\dots, y_p) = - y_{p}g_{p} + h_{p}\,.
\ee

Note that there are now exactly two mode frequencies, $g_{p-1}$ and $g_p$, and there is always an $h_i$ depending on state variables. In order to determine if the system \eqref{yy} satisfies the competitiveness conditions (A)-(D) (and therefore if the original system satisfies these competitiveness conditions), it is sufficient to check if $g_{p-1}=g_p >0$ for some collection of time values. This is only one condition to check, rather than $2^{n-1}-1$ conditions to check from the original system. Therefore, while the conversion of the system \eqref{autonomoussystem} to the equivalent system \eqref{yy} may seem somewhat roundabout, doing so greatly simplifies the search for possible chaotic dynamics under the competitive modes framework.

\section{Discussion}

The construction of Lyapunov functions for nonlinear dynamical systems is often either simple, or quite challenging, with little room in between. Aside from choosing some standard forms (such as the common bowl shape centered about an equilibrium value), there is more an art to the selection of such function. However, as we have demonstrated for the R\"ossler system, it is possible to use differential elimination to obtain a contraction map in a single state variable, which can then be used to obtain a stability result for all state variables. A similar approach was also employed in \cite{MM1} to study Michaelis-Menten enzymatic reactions, and after using a reduced equation, the proof of global asymptotic stability for a positive steady state was rather simple. Differential elimination, and reduction of dimension more generally, can therefore be useful in helping one determine asymptotic properties of solutions to nonlinear dynamical systems. On the other hand, it is now known that there are autonomous chaotic systems which lack any equilibrium \cite{noEQ}. The approach outlined here could still be used to construct maps which demonstrate the boundedness of trajectories in time for such systems, even if such trajectories do not approach any fixed points.

While useful for obtaining contraction maps, reduction of dimension is also a promising tool for finding and studying chaotic trajectories in nonlinear dynamical systems. We were able to show that reduction of dimension can be used to simplify calculations involved in using undetermined coefficient methods \cite{uc1,uc2,uc3,uc4} by a factor of $1/n$, where $n$ is the dimension of the dynamical system, meaning that the number of equations needing to be solved will not increase with the size of the system - i.e. the computational complexity of the approach will not scale with the size of the system but rather will remain fixed. This means that one may approximate chaotic trajectories through such approaches in systems of rather large dimension, without the computational problem becoming unwieldy, provided that the system can be reduced to a single ODE governing only one state variable. 

While chaos is often studied numerically, recently analytical approaches have been employed to construct trajectories approximating chaotic orbits. When employing these methods, it can be beneficial to consider a single equation rather than a system of equations, even if the single equation is more complicated. This is true for series and perturbation approaches, as the reduction of order requires one to track fewer functions, which is particularly useful when dealing with messy equations. Furthermore, analytical approaches permitting the control of error, such as the optimal homotopy analysis method, rely on assigning an error control parameter to each state variable. Reduction of dimension can allow one to minimize error via a single control parameter \cite{MM2}, rather than over multiple parameters \cite{fvk}, which is computationally less demanding. 

The reduction of dimension can also be useful for diagnostic tests for chaos. When applying the competitive modes analysis, a type of diagnostic criteria for finding chaotic in nonlinear dynamical systems, one performs binary comparisons between the mode frequencies of an oscillator corresponding to each equation. However, if one first applies reduction of dimension and reduces the system of a single ODE for one state variable, we prove that only one comparison would needed. This is particularly beneficial when studying large dimensional systems, since the number of naive comparisons needed scales like $2^{n-1}$ in dimension $n$.

In the future, it would be interesting to consider an algorithm that considered elimination not only elements $\pa^n$ ($\pa = \frac{d}{dt}$) of a differential ring $R[[\pa]]$, but also integral operators $\pa^{-n}$ (where $\pa^{-n}$ satisfies $\pa^{-n}\pa^{n} = \pa^0$ and hence is the inversion of the operator $\pa^n$). Indeed, in cases where the differential elimination algorithm failed to give a reduction of the system to a single ODE, we found by manual substitutions that one can arrive at an integro-differential equation. While more complicated, such integro-differential equations can still cast light on the behavior of solutions, and can prove useful in obtaining Lyapunov functions. More generally than for dynamical systems, these inverse operators $\pa^{-n}$ play a role in the study of operators and integrable hierarchies arising in nonlinear evolution PDEs \cite{Mat,solitons,Ma1,Ma2,Ma3,pde1,pde2}. Therefore, the extension of the algorithm to the ring of formal Laurent series in $\pa$ would be a fruitful area for future work, not only for dynamical systems but also for integrable partial differential equations.

\subsection*{Acknowledgements}
HAH gratefully acknowledges the support of EPSRC Fellowship EP/K041096/1.

\appendix 

\section*{Appendix A: List of Chaotic Systems}
When testing our approach, we considered a variety of specific chaotic systems. Results for these are listed in Table 1. As the form and scaling of such equations can vary in the literature, we list the specific form of these equations used in our work.

Let $x,y,z,w\in C^n(\mathbb{R})$ where $n$ is the dimension of the relevant dynamical system, and let $a,b,c,d \in \mathbb{R}$ be parameters. 

Lorenz system \cite{lor1,lor2}:
\be\begin{aligned}
\dot{x} & = a(y-x)\,,\\
\dot{y} & = x(b-z)-y\,,\\
\dot{z} & = xy - cz\,.
\end{aligned}\ee

Modified Chua's circuit \cite{moch1,modch2,modch3}:
\be\begin{aligned}
\dot{x} & = a\left( y - \frac{1}{7}\left(2x^3 -x\right) \right)\,,\\
\dot{y} & = x-y+z\,,\\
\dot{z} & = -by\,.
\end{aligned}\ee

Chen - Lee system \cite{cl1,cl2}:
\be\begin{aligned}
\dot{x} & = ax - yz\,,\\
\dot{y} & = by + xz\,,\\
\dot{z} & = cz + \frac{1}{3}xy\,.
\end{aligned}\ee

Rabinovich - Fabrikant equations \cite{rf1,rf2,rf3}:
\be\begin{aligned}
\dot{x} & = y\left(z-1+x^2\right) + ax\,,\\
\dot{y} & = x\left( 3z+1-x^2 \right) + ay\,,\\
\dot{z} & = -2z(b+xy)\,.
\end{aligned}\ee

R\"ossler system \cite{ros1,ros2}:
\be\begin{aligned}
\dot{x} & = -y-z\,,\\
\dot{y} & = x+ay\,,\\
\dot{z} & = b+z(x-c)\,.
\end{aligned}\ee

Chen system \cite{chen1,chen2}:
\be\begin{aligned}
\dot{x} & = a(y-x)\,,\\
\dot{y} & = (b-a)x-xz+by\,,\\
\dot{z} & = xy-cz\,.
\end{aligned}\ee

L\"u system \cite{lu1,lu2}:
\be\begin{aligned}
\dot{x} & = a(y-x)\,,\\
\dot{y} & = by-xz\,,\\
\dot{z} & = xy - cz\,.
\end{aligned}\ee

T system \cite{t1,t2,t3}:
\be\begin{aligned}
\dot{x} & = a(y-x)\,,\\
\dot{y} & = (b-a)x - axz\,,\\
\dot{z} & = xy - cz\,.
\end{aligned}\ee

4D Qi-Du-Chen-Chen-Yuan system \cite{qii}:
\be\begin{aligned}
\dot{x} & = a(y-x)+yzw\,,\\
\dot{y} & = b(x+y)-xzw\,,\\
\dot{z} & = -cz + xyw\,,\\
\dot{w} & = -dw + xyz\,.
\end{aligned}\ee

Qi-Chen-Du-Chen-Yuan system \cite{qi}:
\be\begin{aligned}
\dot{x} & = a(y-x) + yz\,,\\
\dot{y} & = bx - y - xz\,,\\
\dot{z} & = xy - cz\,.
\end{aligned}\ee

Generalized Lorenz canonical form \cite{genlor1,genlor2,genlor3}:
\be\begin{aligned}
\dot{x} & = ax - (x-y)z\,,\\
\dot{y} & = -by - (x-y)z\,,\\
\dot{z} & = -cz + (x+y)(x+dy)\,.
\end{aligned}\ee

Two-parameter model for the blue-sky catastrophe \cite{bsc1,bsc2,bsc3}:
\be\begin{aligned}
\dot{x} & = \left( 2+a-10\left(x^2 + y^2\right) \right)x + y^2 + 2y + z^2\,,\\
\dot{y} & = -z^3 - (1+y)\left( y^2 + 2y + z^2 \right) - 4x +ay\,,\\
\dot{z} & = (1+y)z^2 + x^2 - b\,.
\end{aligned}\ee
 
4D Lorenz - Stenflo system \cite{ls1,ls2,ls3}:
\be\begin{aligned}
\dot{x} & = a(y-x)+bw\,,\\
\dot{y} & = cx - xz - y\,,\\
\dot{z} & = xy - dz\,,\\
\dot{w} & = -x -aw \,.
\end{aligned}\ee

Genesio-Tesi system \cite{gt1,gt2}:
\be\begin{aligned}
\dot{x} & = y\,,\\
\dot{y} & = z\,,\\
\dot{z} & = ax + by + cz + x^2\,.
\end{aligned}\ee

Arneodo-Coullet-Tresser system \cite{act}:
\be\begin{aligned}
\dot{x} & = y\,,\\
\dot{y} & = z\,,\\
\dot{z} & = ax - by - cz - x^3\,.
\end{aligned}\ee

\section*{Appendix B: List of Hyperchaotic Systems}
When testing our approach, we considered a variety of specific hyperchaotic systems. Results for these are listed in Table 2. As the form and scaling of such equations can vary in the literature, we list the specific form of these equations used in our work. 

Let $x,y,z,w\in C^4(\mathbb{R})$ and let $a,b,c,d,e,f \in \mathbb{R}$ be parameters.

4D R\"ossler flow \cite{hros1}:
\be\begin{aligned}
\dot{x} & = -y-z\,,\\
\dot{y} & = x + 0.25y + w\,,\\
\dot{z} & = 3 + xz\,,\\
\dot{w} & = -0.5z + 0.05w \,.
\end{aligned}\ee

Hyperchaotic Chen system \cite{hchen1,hchen2}:
\be\begin{aligned}
\dot{x} & = a(y-x)\,,\\
\dot{y} & = -bx -xz + cy -w\,,\\
\dot{z} & = xy - dz\,,\\
\dot{w} & = x\,.
\end{aligned}\ee

Hyperchaotic L\"u system \cite{hlu}:
\be\begin{aligned}
\dot{x} & = a(y-x) + w\,,\\
\dot{y} & = by - xz\,,\\
\dot{z} & = xy - cz\,,\\
\dot{w} & = xz + dw\,.
\end{aligned}\ee

Modified hyperchaotic L\"u system \cite{hmodlu}:
\be\begin{aligned}
\dot{x} & = a(y-x+yz)\,,\\
\dot{y} & = by - xz +w\,,\\
\dot{z} & = xy - cz\,,\\
\dot{w} & = -dx\,.
\end{aligned}\ee

Hyperchaotic Wang-Liu system \cite{wl}:
\be\begin{aligned}
\dot{x} & = a(y-x)\,,\\
\dot{y} & = bx -cxz + w\,,\\
\dot{z} & = - dz + ex^2\,,\\
\dot{w} & = -fx\,.
\end{aligned}\ee

Hyperchaotic Jia system \cite{jia1,jia2}:
\be\begin{aligned}
\dot{x} & = a(y-x) + w\,,\\
\dot{y} & = bx -xz - y\,,\\
\dot{z} & = xy - cz\,,\\
\dot{w} & = dw -xz\,.
\end{aligned}\ee

Hyperchaotic Qi - van Wyk - van Wyk - Chen system \cite{qvvc1,qvvc2}:
\be\begin{aligned}
\dot{x} & = a(y-x) + yz\,,\\
\dot{y} & = b(x+y) - xz\,,\\
\dot{z} & = - cz - dw + xy\,,\\
\dot{w} & = ez - fw +xy\,.
\end{aligned}\ee

\end{document}